\newcommand{\CLASSINPUTtoptextmargin}{2.4cm}
\newcommand{\CLASSINPUTbottomtextmargin}{4.6cm}
\documentclass[conference]{IEEEtran}
\IEEEoverridecommandlockouts
\usepackage{cite}
\usepackage[fleqn]{amsmath}
\usepackage{amssymb,amsfonts}
\usepackage{graphicx}
\usepackage{textcomp, caption, url}
\usepackage{color,soul}
\usepackage{booktabs}
\usepackage{breakurl}
\usepackage{algorithm}
\usepackage{algpseudocode}
\usepackage{graphicx} 
\usepackage{multirow} 
\usepackage{tabularx}

\def\BibTeX{{\rm B\kern-.05em{\sc i\kern-.025em b}\kern-.08em
    T\kern-.1667em\lower.7ex\hbox{E}\kern-.125emX}}
\begin{document}

\title{HTTP REST API Structure Learning}

\author{
\IEEEauthorblockN{Ran Dubin, Amit Dvir, Senior Member, IEEE}
\IEEEauthorblockA{\textit{Department of Computer and Software Engineering} \\ \textit{Ariel Cyber Innovation Center}\\ Ariel University, Israel\\
rand, amitdv@ariel.ac.il}
}
\maketitle

\begin{abstract}
Application Programming Interfaces (APIs) are essential in software development, enabling web services, mobile apps, and microservices. However, their widespread use introduces significant security risks, highlighting the importance of API security. This paper presents HTTP REST API Learning (HRAL), a novel unsupervised anomaly detection approach that models the structure and behavior of API endpoints directly from network traffic, without relying on predefined rules or documentation. HRAL enables robust detection of malicious activity by understanding how APIs behave and flagging deviations as potential threats. We evaluate HRAL across varying levels of OpenAPI documentation detail and compare it with existing techniques. HRAL achieves strong performance, with an average recall of 82.07\% and an F1-score of 87.24\%, significantly outperforming when API documentation is limited. Moreover, our results approach the effectiveness of full API document definitions. When combined with signature-based rules such as the OWASP ModSecurity CRS, our system 
achieves 100\% detection when combined with signature-based rules (such as OWASP), while HRAL alone achieves 82.07\% recall — significantly outperforming all documentation-based baselines except Full OpenAPI.
These results highlight HRAL’s effectiveness in real-world, partially documented API environments and its potential as a foundational layer for modern API security solutions.
\end{abstract}


\section{Introduction}
\label{sec:introduction}
As the digital age advances, Application Programming Interfaces (APIs) have become pivotal in software development and interaction, powering everything from web services to mobile apps and microservices architectures. Yet, this prominence comes with substantial risk. API security safeguards the interfaces through which applications communicate, thereby protecting the integrity of the transmitted data and the underlying systems. 

API security involves various procedures and protocols to detect and prevent threats or attacks on API resources to maintain integrity, confidentiality, and availability~\cite{sun2022research}. Moreover, the need for robust API security becomes apparent when we consider the consequences of its absence. APIs are a prominent target for cyber attacks because they provide direct access to sensitive data and services~\cite{mendoza2018mobile}.

A successful attack can have detrimental effects, including data theft, financial loss, and reputational damage. In the NoxPlayer attack~\cite{NoxPlayer_eset}, hackers exploited a vulnerability in the application's update API. They manipulated it to distribute malware to users by disguising it as legitimate updates. The attack underscores the need for robust authentication and server identification in API requests, as the lack of these led to the successful breach. 

Another well-known API vulnerability is Log4j~\cite{CVE-2021-44228, 10037384}. The Log4j API attack in December 2021 exploited a vulnerability in Apache Log4j, allowing attackers to execute malicious code remotely. With a well-crafted request, attackers could manipulate a system to download and run harmful payloads. As a result, numerous systems were scanned and exploited across the internet. OpenAPI~\cite{openapi}, formerly known as Swagger, is a foundational tool in modern software development for describing and documenting RESTful APIs. It provides a standardized, machine-readable specification—typically written in YAML or JSON that outlines API endpoints, request parameters, response formats, and authentication mechanisms. By enabling automated tools and facilitating seamless integration across services, OpenAPI significantly simplifies the development and maintenance of web APIs. However, in practice, OpenAPI documentation is often incomplete, outdated, or entirely missing, posing challenges for security analysis and automated tooling. 

A balance between securing the organization, controlling API access, and providing necessary access is fundamental. While API security is considered a foundational component in the industry, there has been a surprising lack of academic investigation. Existing academic research primarily focuses on detecting HTTP REST API attacks using machine learning techniques~\cite{10.1007/978-3-031-35314-7_33}. However, these detection-based approaches may be prone to false positives, potentially blocking legitimate user requests or tackling specific attack vectors. 

Therefore, generic solutions that can tackle several vector attacks are needed. Consequently, this paper introduces an innovative solution for understanding HTTP REST API endpoint representations, designed to detect anomalous API requests by modeling normal API behavior. The proposed method, HTTP REST API Learning (HRAL), leverages either observed network traffic or existing OpenAPI documentation as its input. Recognizing that API documentation is often missing or only partially complete, we propose an algorithm that enriches or reconstructs the OpenAPI specification derived from observed traffic; in the absence of documentation, it generates a specification from scratch. This capability enhances the completeness and accuracy of API understanding, thereby improving the detection of attacks targeting API endpoints. To summarize, this work evaluates two unsupervised API learning methods, HRAL and Speculator ~\cite{SpeculatorOpenClarity}. Both methods are compared against three supervised approaches that leverage OpenAPI documentation at varying levels of completeness. 

The remainder of this paper is structured as follows:
The works related to our study are presented in Section \ref{Related_Works}. Our evaluation and validation methodology is discussed in Section \ref{Methodology}. Section \ref{Dataset} describes our dataset. The results of our HTTP REST API learning are evaluated in Section \ref{Evaluation}. In Section \ref{Limitations}, we identify the limitations of HRAL and suggest areas for future work. Finally, Section \ref{Conclusions} summarizes our conclusions and suggests potential directions for further research.

\section{Related Work}
\label{Related_Works}
Existing academic research primarily focuses on detecting HTTP REST API attacks using machine learning techniques~\cite{ ron2016analysis, hu2021seapp}. However, most of them tackle specific attacks and do not try to detect anomalies, which can be a new zero-day attack.

Baye et al. \cite{9615638} propose an ML-based technique that uses features such as bandwidth and the number of requests per token to detect and classify API traffic. The authors compare their method with the well-known rule-based IDSs Snort and Suricata. Alam et al. \cite{9720840} investigate and analyze the API implementation risks in the Core Banking System and recommend the best practices to secure API implementation.  Espunyes et al. \cite{sanchez2022machine} proposed API detection based on an ensemble of Sentence Bert with unsupervised DBSCAN and achieved an F1-score of 0.9896. Moradi et al. \cite{moradi2019auto} introduced an unsupervised anomaly detection technique utilizing an Auto-Encoder LSTM for feature extraction and an Isolation Forest as a classifier. They applied this model to the CSIC 2010 dataset, achieving an F1 Score of 81.96\%. 

In more recent research \cite{moradi2021auto}, the authors proposed an unsupervised Deep Support Vector method and compared two feature extraction strategies—one-hot and bigram — using raw requests. They evaluated the model on the CSIC-2010 and ECML/PKDD-2007 datasets, achieving F1 Scores of 89\% and 79.48\%, respectively. 

Arstila et al. \cite{leila2023securing} also proposed an unsupervised LSTM Autoencoder to detect anomalies in microservice gRPC (Google Remote Procedure Call) logs. However, our work focuses on understanding the API structure to improve the detection of abnormal behavior. Moreover, to our knowledge, the API structure was not evaluated. Therefore, our goal is to show the effectiveness of the API structure for detecting abnormal behavior. Paul \cite{Paul24} examined the significance of anomaly detection in the context of API security, detailing
various machine learning methods. The author explored how these approaches facilitate real-time monitoring, dynamic threat detection, and the complexities of establishing baseline behavior in diverse API environments.

The Speculator~\cite{SpeculatorOpenClarity} is an open-source library designed to reconstruct OpenAPI specifications from HTTP traffic automatically. It accomplishes this by dissecting each incoming request and transforming its constituent elements into structured objects. These objects encompass a wide range of request components, including the request path, headers, query parameters, and schema details. By capturing and organizing this information, the Speculator intelligently identifies API endpoints and assembles them into a coherent specification. The Speculator produces a crafted, OpenAPI-compliant specification that provides a comprehensive view of how the API is used. This specification outlines the exposed endpoints and delineates the data formats employed, facilitating a thorough understanding of the API's behavior and functionality. As this work is the most closely aligned with our research on structuring API requests to OpenAPI, we will adopt it as our baseline.

\section{Methodology}
\label{Methodology}
Our approach centers on modeling the API structure to detect anomalies that may indicate attacks targeting HTTP REST API endpoints. Effectively learning the API endpoint structure requires a thorough understanding of the API's expected behavior.

Given that OpenAPI documentation is often incomplete or unavailable in real-world deployments, our system infers the API structure directly from observed API requests, constructing a behavioral model based on actual usage. This requires monitoring live traffic to accurately reconstruct the API's intended behavior. By extracting the structural patterns from real requests and expressing them in OpenAPI format, our method enhances and enriches API endpoint descriptions, mitigating the limitations inherent in existing developer documentation.
Therefore, in Section~\ref{Supervised-HRAL}, we first explain the OpenAPI documentation-level completeness, and then, in Section~\ref{unsupervised-method}, two unsupervised methods, \textbf{HRAL} and \textbf{Speculator}, are presented. 
\begin{table*}[!]
\begin{tabular}{|p{5cm}|p{5cm}|p{5.5cm}|}
  \textbf{Minimal} &
  \textbf{Basic} &
  \textbf{Full} \\
   \begin{tabular}[c]{@{}l@{}}"/bookstore/\{username\}/account\_info": 
   \{\\     "get": 
    \{\\         "summary": "Book Store Account Info",\\         "operationId": "book\_store\_account\_info  \\ \_bookstore\_username\_account\_info\_get",\\         "parameters": 
    {
    [
    }\\             
    \{\\                 "required": true,\\                 "schema": 
        \{\\                     "title": "Username",\\                     "type": "string"\\                 
        \},\\                 "name": "username",\\                 "in": "path"\\             
    \},\\             
    \{\\                 "required": true,\\                 "schema": 
        \{\\                     "title": "Api Key",\\ \\                     "type": "string"\\       \},\\                 "name": "api\_key",\\                 "in": "query"\\             
    \}\\         {]}\\     \}\\\}
    \end{tabular} &
  
  \begin{tabular}[c]{@{}l@{}}"/bookstore/\{username\}/account\_info": 
  \{\\     "get": 
    \{\\         "summary": "Book Store Account Info",\\         "operationId": "book\_store\_account\_info\\\_bookstore\_username\_account\_info\_get",\\         "parameters": 
    {
    [
    }\\             
    \{\\                 "required": true,\\                 "schema": 
        \{\\                     "title": "Username",\\                     "pattern": "\textasciicircum{}{[} A-Z a-z {]}{[} A-Z a-z 0-9 {]}$*$",\\   "type": "string"\\          \},\\                 "name": "username",\\                 "in": "path"\\
    \},\\            
    \{,\\                 "required": true,\\                 "schema": 
        \{\\                     "title": "Api Key",\\                     "pattern": "\textasciicircum{}{[} A-Z a-z {]}{[} A-Z a-z 0-9 {]}$+$",\\                     "type": "string"\\                 
        \},\\                 "name": "api\_key",\\                 "in": "query"\\             
        \}\\         {]}\\     \}\\\}
  \end{tabular} &
  
  \begin{tabular}[c]{@{}l@{}}"/bookstore/\{username\}/account\_info": 
  \{\\     "get": 
    \{\\         "summary": "Book Store Account Info",\\         "operationId": "book\_store\_account\\\_info\_bookstore\_username\_account\_info\_get",\\         "parameters": 
    {
    [
    }\\             
    \{\\                 "required": true,\\                 "schema": 
        \{\\                     "title": "Username",\\                     "maxLength": 20,\\                     "minLength": 1,\\                     "pattern": "\textasciicircum{}{[} A-Z a-z {]}{[} A-Z a-z 0-9 {]}$*$",\\                     "type": "string"\\                 
        \},\\                 "name": "username",\\                 "in": "path"\\             
        \},\\             
        \{\\                 "required": true,\\                 "schema": 
            \{\\                     "title": "Api Key",\\                     "maxLength": 43,\\                     "minLength": 1,\\                     "pattern": "\textasciicircum{}{[} A-Z a-z {]}{[} A-Z a-z 0-9 {]}$+$",\\                     "type": "string"\\                \},\\                 "name": "api\_key",\\                 "in": "query"\\             \}\\         {]}\\     \}\\\}
    \end{tabular}
 \end{tabular}
\caption{Comparison of three tiers of OpenAPI documentation, each serving as an input for the HRAL system. The objective is to emphasize the varying degrees of detail across these tiers. A richer level of detail not only aids in better understanding the API's structure but also strengthens the system's security.}
\label{tab:OpenAPI_doc}
\end{table*}

\subsection{OpenAPI Documentation-Based}  
\label{Supervised-HRAL}
We define three levels of OpenAPI documentation, and provide examples in Table \ref{tab:OpenAPI_doc}.
Represents users unfamiliar with OpenAPI security practices and provides limited documentation. The schema mainly specifies parameters with their general types, such as \texttt{string} and \texttt{boolean}. textbf{Basic}: Corresponds to the most commonly observed level of detail in OpenAPI specifications found online. Beyond the \textit{Minimal} level, it incorporates basic regex-based pattern constraints for expected input values. \textbf{Full}: Offers a comprehensive level of detail, often exceeding typical developer-provided documentation. It extends the \textit{Basic} level by specifying minimum and maximum field lengths and enforcing stricter regex patterns. The OpenAPI documentation-based approach serves as the ground truth for evaluating our proposed methods, as it closely reflects the specification standards used in real-world API deployments.

\subsection{Unsupervised HRAL}
\label{unsupervised-method}
HRAL request-based approach aims to attain the granularity of the \textit{Full} documentation tier, as it offers a robust defense against abnormal structure attacks. As mentioned above, the OpenAPI specification might sometimes be incomplete (e.g., Minimal, Basic) or unavailable. We, therefore, introduced an OpenAPI request-based algorithm that harnesses observed traffic to understand the API behavior.

\begin{figure*}[!]
    \centering
\includegraphics[width=16 cm, height=10 cm]{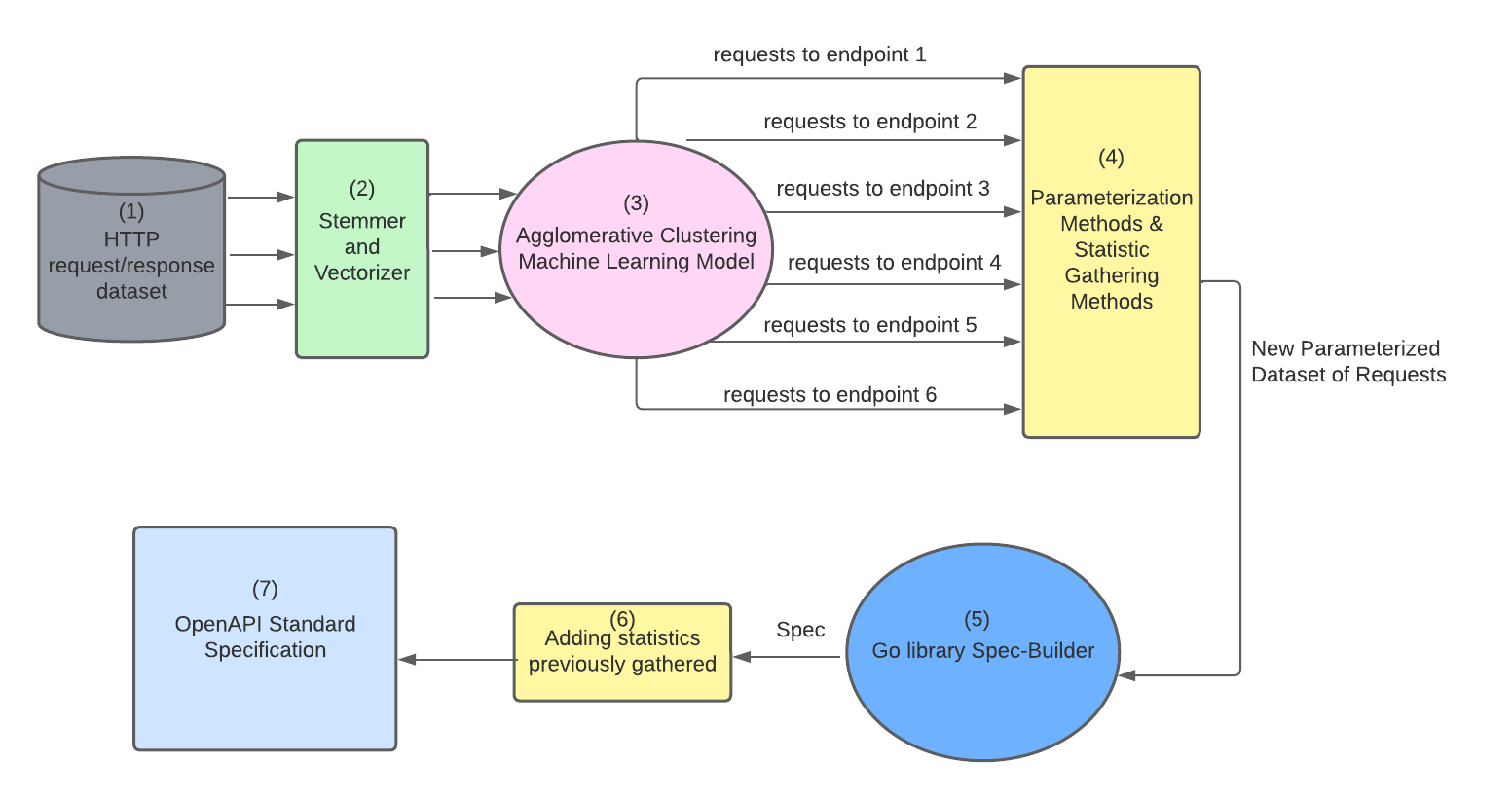}
    \caption{OpenAPI Request-Based Algorithm}
    \label{fig:unsupervised_learning}
\end{figure*}
Figure \ref{fig:unsupervised_learning} shows HRAL architecture. Given an HTTP request or Response, Step (1), the system (at Step (2)) processed them using the Porter Stemmer \cite{porter1980algorithm} and then vectorized via Count Vectorizer. Subsequently, they are channeled into an Agglomerative Clustering model~\cite{monath2021scalable}, which groups them by API endpoint type, as illustrated in Step (3). Initially, all requests are treated as a single cluster, but the algorithm restructures them to ensure intragroup similarity and intergroup distinction, using the Ward linkage~\cite{murtagh2014ward} with a distance threshold of 2. Our technique stands out for its ability to determine the optimal cluster count and adapt to APIs with variable endpoint counts. Within each cluster, paths are juxtaposed to discern regularities (endpoints) and irregularities (path parameters). The threshold value of 2 was selected empirically by evaluating clustering quality across thresholds ranging from 1 to 5, using silhouette score as the optimization criterion on a held-out subset of the training traffic. 

This parameterization process, depicted in Step (4), differentiates between static and variable values in each path. While aggregating these parameters, we also collate statistics to enrich the final specification, including parameter lengths, types, and Regular Expression patterns. Afterwards, the data set, enriched with these parameters, feeds into the Speculator ~\cite{SpeculatorOpenClarity} for further processing (Steps 5-7). 

\begin{figure*}[!]
    \centering
\includegraphics[width=0.7\textwidth]{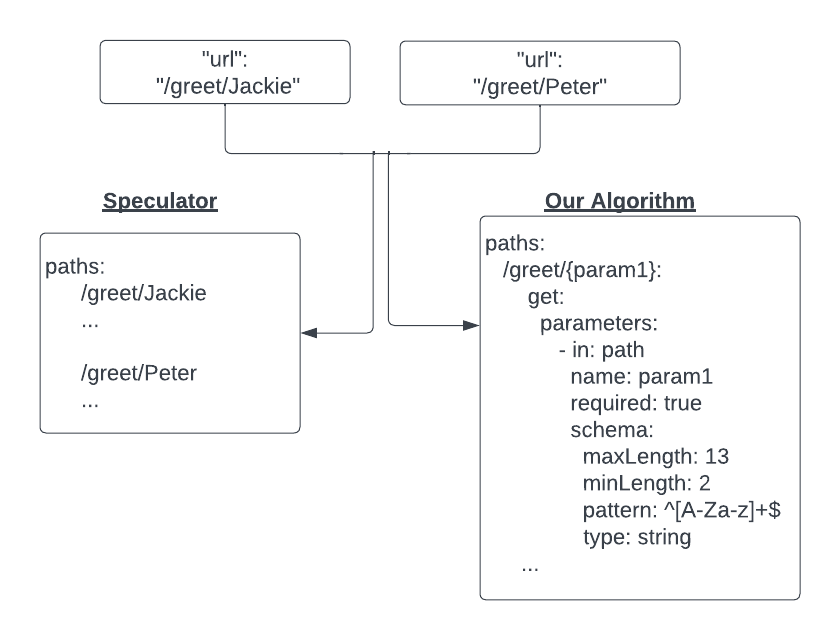}
    \caption{OpenAPI Request-based and Spec-building Algorithm}    \label{fig:parameterization}
\end{figure*}

Using the Speculator ~\cite{SpeculatorOpenClarity} as a baseline tool, we have made pivotal enhancements to address threats more effectively. Our enhancements include a refined clustering mechanism and an innovative method for discerning string-based path parameters before introducing them to the Speculator. The original Speculator struggles to discern string-based path parameters, often resulting in less accurate specifications. By augmenting its parameter criteria, we have transcended this limitation. The example in Fig. \ref{fig:parameterization} shows that the Speculator might interpret two distinct requests to an endpoint as separate paths, whereas our algorithm can detect path parameters and document them appropriately. 

Since our algorithm clusters the requests by API endpoint before feeding the data to the Speculator, we learn valuable insights into which parts of the path are parameters and which are endpoints. After identifying path parameters, our algorithm gathers statistics on each parameter and detects abnormal behavior. To further elaborate on our contribution, we provide the following pseudo-code for both the Speculator algorithm and our proposed algorithm.
\begin{algorithm}[!]
  \caption{Speculator Algorithm}\label{Spec_Algo}
  \begin{algorithmic}[1]
     \State load all requests
     \For{for each request} 
        \State Creates a new endpoint for this path if it does not yet exist
        \State Extract query parameters from the request and add them to the endpoint
        \State Extract headers from the request and response, and add to the endpoint
        \State Add an endpoint to the spec.
    \EndFor
  \end{algorithmic}
\end{algorithm}
\begin{algorithm}[!]
  \caption{Our OpenAPI Request-Based Algorithm}\label{Our_Algo}
  \begin{algorithmic}[1]
     \State load all requests
     \For{for each cluster} 
        \State Discover the path parameters from the requests in the cluster
        \State Parameterize the path of each endpoint in the cluster
    \EndFor
        \State Extract stats for each endpoint
        \State Runs the speculator algorithm on all of the requests
        \State Load the built spec
        \State Add the stats for each endpoint to the spec
        \State Update the spec with the parameterized paths
  \end{algorithmic}
\end{algorithm}
\section{Dataset}
\label{Dataset}
The well-known dataset for API security is  CSIC 2010 ~\cite{CSIC2010}. Unfortunately, CSIC 2010 is outdated and lacks the API baseline needed to construct unsupervised and supervised API structure learning for modern cybersecurity challenges. As a result, we are using Aharon et al.  \cite{aharon2025classification}, which created an API security dataset with a baseline and relevant API-challenging attacks, named the API Traffic Research Dataset Framework (ATRDF)~\cite{aharon2025classification} as can be seen in Table \ref{table:attack_count}. 

To our knowledge, ATRDF is the only publicly available dataset that provides both labeled API endpoint structure and realistic attack traffic, making cross-dataset validation currently infeasible. ATRDF includes 18 API endpoints and four attack vectors with fully automated randomization. The dataset consists of attacks such as Directory Traversal, Cookie Injection, LOG4J ~\cite{Log4JCVE} (11,721 samples), Remote Code Execution (RCE, 12,293 samples), Log Forging~\cite{owsap-log-forge}(12,059 samples), SQL Injection (SQLi, 24,285 samples)~\cite{owsap-SQL-injection}, and XSS (12,403 samples). The dataset contains 108,252 normal and 109,277 abnormal sets of requests and responses. 

. We encourage the community to release additional API security datasets to enable broader evaluation in future work

\begin{table}[!]
\begin{center}
\begin{tabular}{|p{4cm}|r|}
\toprule
Attack Type         & Count \\
\midrule
Cookie Injection    & 24,458 \\
Directory Traversal & 12,058 \\
Log Forging         & 12,059 \\
LOG4J               & 11,721 \\
RCE                 & 12,293 \\
SQL Injection       & 24,285 \\
XSS                 & 12,403 \\%
\bottomrule
\end{tabular}

\caption{Counts of Each Attack Type on ATRDF Dataset}
\label{table:attack_count}
\end{center}%
\end{table}

\section{Evaluation}
\label{Evaluation}
The evaluation of this study focuses on the crucial task of detecting and identifying anomalous and malicious HTTP requests in a dataset of API traffic by learning the API endpoint representation. By detecting abnormal patterns in API requests, we can identify changes to API endpoints, new APIs, or new attacks. In our evaluation, we have the following abnormal detection behaviors: \textbf{Minimal/Basic/Full OpenAPI (Supervised) spec}: The ground truth evaluation method does not utilize API understanding; instead, it relies solely on a comparison with the OpenAPI specification, which can come in three levels of information depth. \textbf{Speculator spec}: Evaluation based on the Speculator unsupervised algorithm presented in Algorithm 1. \ref{Spec_Algo} is constructed from the incoming traffic flow. \textbf{Unsupervised spec(HRAL)}: Evaluate using our custom-built Unsupervised API Specification Algorithm, constructed from incoming traffic, which enables a more thorough examination of potential anomalies and malicious requests as described in Algorithm \ref{Our_Algo}.
   
Evaluation incorporates specification anomalies, further enhancing our ability to identify and mitigate potentially harmful traffic. This multifaceted evaluation framework aims to provide a robust assessment of the effectiveness and reliability of the proposed detection techniques.

First, we evaluate detection capabilities based on three levels of supervised API documentation: Minimal, Basic, and Full. These levels reflect typical real-world scenarios ranging from incomplete to comprehensive documentation. In this evaluation, anomaly detection is based solely on the structural knowledge derived from the HTTP REST API documentation, without applying explicit detection rules.

As shown in Table~\ref{table:3}, the Minimal OpenAPI documentation yields poor detection performance, with an average recall of only 22.94\%, failing to detect most attack types. The Basic OpenAPI configuration significantly improves detection capabilities, achieving an average recall of 72.30\%, but still struggles to detect several attack types such as Cookie Injection, Directory Traversal, Log4j, and RCE. With Full OpenAPI documentation, the system reaches an average recall of 93.57\%, demonstrating near-complete detection coverage—yet even in this ideal setting, Log4j attacks remain relatively harder to detect with a recall of only 40.09\% this is due to the fact that not all the attacks are found in the API endpoint and some are also are found in the request body and this solution is not designed to cover that. As a result, there is a need for more extended detection ~\cite{aharon2025classification}. The Speculator approach performs slightly better than Minimal OpenAPI with an average recall of 23.84\%, and it successfully detects Directory Traversal, RCE, and Log4j. However, it completely fails to detect Cookie Injection, Log Forging, SQL Injection, and XSS, highlighting limitations in its general applicability. In contrast, HRAL shows robust detection performance across all attack types, achieving an average recall of 82.07\% and an F1-score of 87.24\%. HRAL is second only to the Full OpenAPI configuration, but with the key advantage that it does not require documentation. It significantly outperforms both the Minimal and Basic OpenAPI modes, as well as the Speculator method, making it a highly practical and effective solution in environments where API documentation is incomplete, outdated, or unavailable. This demonstrates that HRAL can greatly enhance API security in real-world scenarios, where ideal documentation is rarely available.

To address the limitations of body-based attack detection, which were previously underrepresented, we incorporated signature-based detection rules. The results, summarized in Table~\ref{table:4}, demonstrate the effectiveness of our proposed algorithm, achieving a perfect detection rate of 100\% for malicious requests. This outcome is particularly significant, as certain attack vectors are designed to adhere to the structural constraints of APIs, thereby bypassing anomaly-based detection systems. In such scenarios, signature-based methods—such as those outlined in the OWASP ModSecurity Core Rule Set~\cite{OWASP-ModSecurity-CRS}—play a critical role in closing the detection gap left by unsupervised techniques. It is important to note that these findings are based on a controlled baseline evaluation.

\begin{table*}
\centering
\begin{tabular}{|p{4.1cm}|p{4.1cm}|c|c|}
\toprule
\textbf{Attack Type} & \textbf{Method} & \textbf{Recall [\%]} & \textbf{F1-score [\%]} \\
\midrule
\multirow{5}{*}{Cookie Injection}
  & Minimal OpenAPI   & 100 & 100 \\
  & Basic OpenAPI     & 100 & 100 \\
  & Full OpenAPI      & 100 & 100 \\
  & HRAL              & 99.48 & 99.74 \\
  & Speculator   & 0   & 0 \\
\midrule
\multirow{5}{*}{Directory Traversal}
  & Minimal OpenAPI   & 0    & 0 \\
  & Basic OpenAPI     & 29.54 & 45.61 \\
  & Full OpenAPI      & 100 & 100 \\
  & HRAL              & 99.5 & 99.97 \\
  & Speculator   & 100 & 100 \\
\midrule
\multirow{5}{*}{LOG4j}
  & Minimal OpenAPI   & 5.17 & 9.83 \\
  & Basic OpenAPI     & 19.11 & 32.09 \\
  & Full OpenAPI      & 40.09 & 57.24 \\
  & HRAL              & 37.17 & 54.18 \\
  & Speculator   & 14.54 & 25.39 \\
\midrule
\multirow{5}{*}{Log Forging}
  & Minimal OpenAPI   & 0 & 0 \\
  & Basic OpenAPI     & 100 & 100 \\
  & Full OpenAPI      & 100 & 100 \\
  & HRAL              & 44.14 & 58.59 \\
  & Speculator   & 0 & 0 \\
\midrule
\multirow{5}{*}{RCE}
  & Minimal OpenAPI   & 0 & 0 \\
  & Basic OpenAPI     & 0 & 0 \\
  & Full OpenAPI      & 100 & 100 \\
  & HRAL              & 96.90 & 99.43 \\
  & Speculator   & 100 & 100 \\
\midrule
\multirow{5}{*}{SQL Injection}
  & Minimal OpenAPI   & 0 & 0 \\
  & Basic OpenAPI     & 100 & 100 \\
  & Full OpenAPI      & 100 & 100 \\
  & HRAL              & 99.60 & 99.80 \\
  & Speculator   & 0 & 0 \\
\midrule
\multirow{5}{*}{XSS}
  & Minimal OpenAPI   & 0 & 0 \\
  & Basic OpenAPI     & 100 & 100 \\
  & Full OpenAPI      & 100 & 100 \\
  & HRAL              & 100 & 100 \\
  & Speculator   & 0 & 0 \\
\midrule
\multirow{5}{*}{Summary}
  & Accuracy (Minimal) & 22.94 & 22.94 \\
  & Accuracy (Basic)   & 72.30 & 72.30 \\
  & Accuracy (Full)    & 93.57 & 93.57 \\
  & Accuracy (HRAL)    & 82.07 & 87.24 \\
  & Accuracy (Speculator)    & 23.84 & 23.84 \\
\bottomrule
\end{tabular}
\caption{Classification Report for Baselines, Speculator and HRAL Without Detection (Precision column removed)}
\label{table:3}
\end{table*}

\begin{table*}[htb]
\centering
\renewcommand{\arraystretch}{1.1}
\setlength{\tabcolsep}{3pt} 
\begin{tabular}{|l|cc|cc|cc|cc|cc|}
\hline
\textbf{Attack} & 
\multicolumn{2}{c|}{Min} & 
\multicolumn{2}{c|}{Basic} & 
\multicolumn{2}{c|}{Full} & 
\multicolumn{2}{c|}{HRAL} & 
\multicolumn{2}{c|}{Spec} \\
& Rec & F1 & Rec & F1 & Rec & F1 & Rec & F1 & Rec & F1 \\
\hline
Cookie Inj.       & 100  & 100  & 100  & 100  & 100  & 100  & 100  & 100  & 100  & 100  \\
Dir. Trav.        & 100  & 100  & 100  & 100  & 100  & 100  & 100  & 100  & 100  & 100  \\
LOG4j             & 100  & 100  & 100  & 100  & 100  & 100  & 100  & 100  & 100  & 100  \\
Log Forging       & 0.06 & 0.12 & 100  & 100  & 100  & 100  & 100  & 100  & 0.06 & 0.12 \\
RCE               & 60.2 & 75.2  & 60.2 & 75.2 & 100  & 100  & 100  & 100  & 100  & 100  \\
SQL Inj.          & 100  & 100  & 100  & 100  & 100  & 100  & 100  & 100  & 100  & 100  \\
XSS               & 100  & 100  & 100  & 100  & 100  & 100  & 100  & 100  & 100  & 100  \\
\hline
Accuracy          & 84.5 & 84.5 & 95.5 & 95.5 & 100  & 100  & 100  & 100  & 89.0 & 89.0 \\
Macro Avg         & 70.0 & 71.9 & 82.5 & 84.4 & 100  & 100  & 100  & 100  & 75.0 & 75.0 \\
Weighted Avg      & 84.5 & 86.2 & 95.5 & 97.2 & 100  & 100  & 100  & 100  & 89.0 & 89.0 \\
\hline
\end{tabular}
\caption{Detection performance comparison powered by detection rules (Recall and F1 in \%).}
\label{table:4}
\end{table*}

\section{Limitations}
\label{Limitations}
While the HRAL system performs well at detecting API-based attacks, it has several limitations due to its clustering-based framework. First, scalability and performance are significant concerns. Clustering algorithms, especially those applied to high-dimensional representations of API calls, are computationally intensive and may not scale efficiently in real-time or large-scale environments. 

Second, HRAL is less adaptable to evolving APIs. Structural changes in the API, such as adding endpoints or modifying schemas, can significantly affect the clustering process's accuracy. Adapting to these changes typically requires retraining or re-clustering, which can introduce operational complexity and latency. 

Third, HRAL lacks interpretability and explainability. Because the model operates unsupervised and does not produce human-readable rules, it provides security analysts with limited insight into why a specific request was flagged as anomalous. This limitation hampers root-cause analysis, incident response, and effective remediation efforts. 

Finally, HRAL does not inherently support anomaly explainability, making it challenging to communicate or visualize how and where anomalies deviate from normal behavior within the API structure. 

Future work should focus on addressing these limitations by incorporating more explainable and adaptive techniques, potentially by hybridizing clustering with rule-based or interpretable machine learning methods.

\section{Conclusion}
\label{Conclusions}
This work introduces an approach for detecting anomalies in HTTP REST APIs by learning the structure and behavior of API endpoints. API endpoints visibility is a key for enforcing security, and above that, detection paradigmas can be enforced \cite{aharon2025classification}. This work focuses on creating an API endpoint structure, learning, and validating it against the current state-of-the-art API structure learning. We evaluate detection using three levels of OpenAPI documentation: Minimal, Basic, and Full, which shows that detection performance improves with documentation quality. 

However, even Full documentation misses some attack types, such as Log4j, because they occur in request bodies rather than endpoints. To address this, we propose HRAL, a documentation-free anomaly detection method that learns endpoint behavior from traffic patterns. HRAL achieves high detection performance (82.07\% recall, 87.24\% F1-score), significantly outperforming Minimal and Basic documentation settings. It is especially useful in real-world environments where API documentation is incomplete or unavailable. 

To improve detection of payload-based attacks, we combine HRAL with signature-based rules (e.g., OWASP ModSecurity CRS), achieving 100\% detection in a controlled setting. This hybrid approach ensures full coverage by detecting both structural and content-level anomalies.  We used ATRDF dataset since it is the only open-source dataset that can support API endpoint learning.

Overall, our findings highlight that modeling REST API endpoint behavior is a powerful and practical method for anomaly detection, and HRAL offers an effective solution for securing APIs in dynamic or undocumented environments.

\bibliographystyle{IEEEtran}  
\bibliography{ref.bib}

\end{document}